# The nature of transport and ferromagnetic properties of the GaAs structures with the Mn δ-doped layer

A.V. Kudrin[*], O.V. Vikhrova, Yu.A. Danilov, M.V. Dorokhin, I.L. Kalentyeva, A.A. Konakov, V.K. Vasiliev, D.A. Pavlov, Yu.V. Usov and B.N. Zvonkov

*Lobachevsky State University of Nizhny Novgorod, Gagarina pr. 23/3, 603950, Nizhny Novgorod, Russia*
*kudrin@nifti.unn.ru

We investigate the nature of transport and ferromagnetic properties of the epitaxial GaAs structure with the Mn δ-doped layer. To modify the properties of the structure electrically active radiation defects are created by irradiation with 50 keV helium ions and a fluence in the range of $1 \times 10^{11}$ - $1 \times 10^{13}$ cm$^{-2}$. The investigations show that transport properties of the structure are determined by two parallel conduction channels (the channel associated with hole transport in a valence band and the channel associated with electron transport in the Mn-related impurity band) and that ferromagnetic properties are determined by electrons localized at allowed states within the Mn impurity band. The ferromagnetic properties of the Mn δ-layer region cannot be affected by the closely located InGaAs quantum well, since the presence of quantum well has negligible influence on the Mn impurity band filling by electrons.

## 1. Introduction

Gallium arsenide heavily doped with Mn is the most studied magnetic semiconductor in recent decades. The nature of ferromagnetism in this material is the subject of discussion despite the enormous amount of experimental and theoretical studies that were carried out. Traditionally two models of ferromagnetism in (Ga,Mn)As are considered: Zener's *p-d* exchange model, which relates the ferromagnetism to the exchange interaction between Mn atoms via carriers in a valence band (holes) [1], and Zener's double exchange model, in which the exchange interaction is realized via carriers in the Mn impurity band (IB) separated from the valence band (VB) [2]. It is known that an irradiation with light ions of medium energies (for example, He$^+$ ions with the energy of 50 keV) controllably creates point defects in single-crystal semiconductors. In GaAs structures the main role in the ion-induced change of electrical properties is played by antisites and vacancies introduced in this way which form deep donor and acceptor levels near the middle of a band gap [3,4]. The creation of radiation defects in (Ga,Mn)As by the ion irradiation makes it possible to vary a carrier concentration and to analyze a carrier concentration influence on the magnetic properties [4–9]. The modulation of the magnetic and transport properties of (Ga,Mn)As after ion irradiation was interpreted both within the model considering the determining role of carriers in the valence band [6–9] and within the model taking into acount the determining role of the separated Mn impurity band [4,5]. The general tendency was the weakening of the ferromagnetic properties after irradiation even with minimum ion fluences.

One of the variants of realization of ferromagnetic (Ga,Mn)As is the introduction of Mn impurity in the form of a δ-doped layer. The first works about Mn delta-doped GaAs structures, fabricated by molecular beam epitaxy (MBE), showed that structures with a single Mn δ-layer are not ferromagnetic and for the appearance of ferromagnetism the quantum well (QW) of GaAs/AlGas heterojunction must be located near the delta layer [10,11]. Later we showed that structures with a single Mn δ-layer, formed by the combined method of metal-organic chemical vapor deposition (MOCVD) and pulsed laser deposition (PLD), demonstrate ferromagnetic properties in transport and optical phenomena up to ~ 30 K [12–15]. Here we present the study of the ion irradiation influence of the Mn δ-doped GaAs structure on its transport and magnetotransport properties. The results reveal the nature of ferromagnetism in structures with the single Mn δ-layer and help to understand the features of luminescent diode structures with the Mn δ-layer nearby InGaAs/GaAs [16–20] or GaAsSb/GaAs [21,22] quantum well (QW).

## 2. Experiments

The structure was grown by the combined method of MOCVD and PLD. Firstly, GaAs undoped buffer layer with thickness of ≈ 200 nm was formed on *i*-GaAs (001) substrate at 600 ºC. Then the substrate temperature was decreased to 400 ºC and the Mn δ-layer was formed by laser sputtering of metallic Mn target for 12 seconds in the flux of hydrogen and arsine. Then cap layer with thickness of ≈ 15 nm was formed by the laser sputtering of undoped GaAs target. The structure was irradiated at room temperature with 50 keV He$^+$ ions by the fluence in the range of $1 \times 10^{11}$ - $1 \times 10^{13}$ cm$^{-2}$. To prevent a channeling, the implantation was performed at the angle of about 10º off the normal to a wafer surface and the azimuthal angle of 45º. For 50 keV He$^+$ ions the peak of defects distribution in GaAs host is located at a depth of about 250 nm. To analyze the transport and magnetic properties of the structure dc magnetotransport measurements were carried out in a van der Pauw geometry. Structural properties were investigated by high-resolution cross-sectional transmission electron microscopy (TEM).

## 3. Results

### 3.1. Structural properties

Figure 1(a) shows a high-resolution cross-sectional transmission electron microscopy (HRTEM) image of the structure. The image demonstrates the high crystalline quality of the buffer layer, Mn-doped region and the cap layer. The image does not reveal any second-phase inclusions. The HRTEM image allows us to obtain a fast Fourier transform (FFT) diffraction pattern (the right part of Fig. 1(a)). The FFT diffraction pattern contains diffraction spots corresponding to zinc-blende type lattices of GaAs epitaxial layers and does not contain additional spots from any other crystalline phase. Figure 1(b) shows cross-sectional scanning electron microscopy (SEM) image of the structure. The SEM image contains the clear change of contrast in the Mn-doped region. The intensity profile obtained from the SEM image allows us to conclude that the Mn impurity is localized in ~ 7–8 nm thick region (Fig. 3(c)).

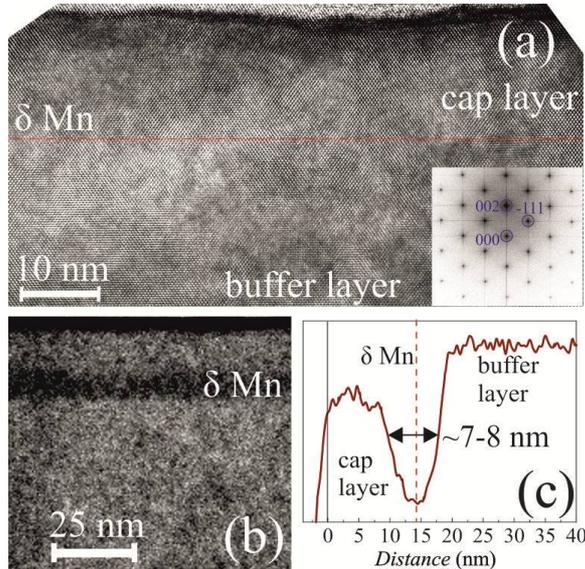

**Fig. 1.** (a) HRTEM image of a 36 × 71 nm² region (at the right, a FFT diffraction pattern of the image). (b) SEM image of the structure. (c) The intensity profile obtained from the SEM image.

### 3.2. Transport properties

Figure 2 exhibits the sheet resistance of the structure as a function of temperature ($R_S(T)$) for different fluences: $F = 0$ (not irradiated), $1 \times 10^{11}$, $1 \times 10^{12}$, $3 \times 10^{12}$ and $1 \times 10^{13}$ cm$^{-2}$. Without irradiation the sheet resistance relatively weakly depends on a temperature in the range of 300 - 90 K and increases with further decreasing temperature. At $T$ about the Curie temperature ($T_C \approx 30$ K) a typical hump is observed. The $R_S(T)$ dependences after irradiation with the fluence of $1 \times 10^{11}$ and $1 \times 10^{12}$ cm$^{-2}$ reveal closer to "metallic" behavior. The sheet resistance shows less modification with decreasing temperature from 295 to 14 K than for the structure before irradiation (Fig. 2). After irradiation with the fluence of $3 \times 10^{12}$ and $1 \times 10^{13}$ cm$^{-2}$ the $R_S(T)$ dependences are semiconductor-like. It will be shown below that the transport properties of the structure are determined by the combination of transport effects related both with carriers energetically localized in the Mn impurity band (IB carriers) and with carriers (holes) in the valence band (VB carriers).

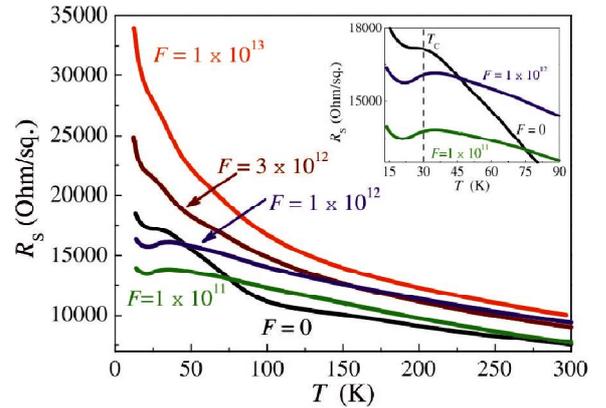

**Fig. 2.** Sheet resistance vs. temperature of the structure before ($F=0$) and after irradiation with different He$^+$ fluences $F$. The inset shows $R_S(T)$ curves for the fluence of 0, $1 \times 10^{11}$ and $1 \times 10^{12}$ cm$^{-2}$.

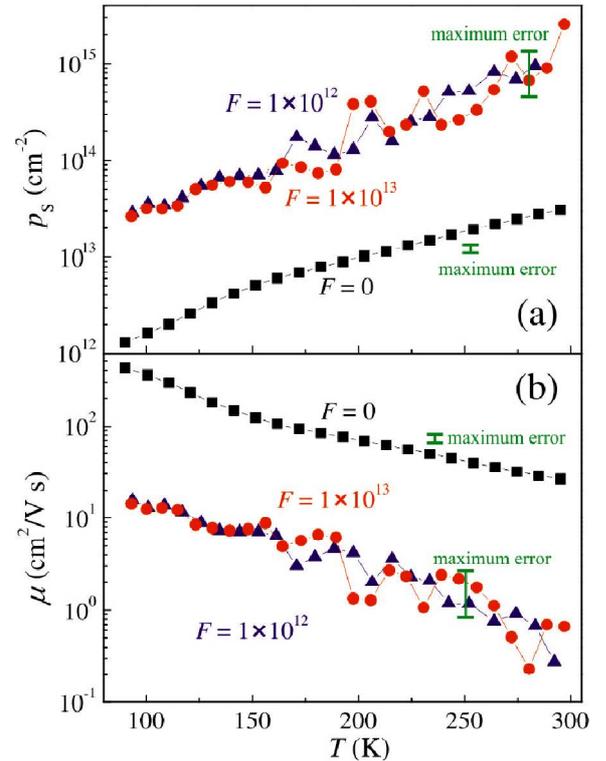

**Fig. 3.** Carrier concentration and mobility vs. temperature of the structure before ($F = 0$) and after irradiation with the fluence of $1 \times 10^{12}$ and $1 \times 10^{13}$ cm$^{-2}$.

Figure 3 demonstrates the values of carrier concentration and mobility in the temperature range of 90 - 295 K for the initial structure and after irradiation with the fluence of $1 \times 10^{12}$ and $1 \times 10^{13}$ cm$^{-2}$. The Hall coefficients were obtained in the constant magnetic field of 0.517 T under the assumption that the ordinary Hall effect predominates at temperatures above 90 K (that is based on a relatively low $T_C \approx 30$ K). After irradiation the significant (by more than an order of magnitude) concentration increase and mobility decrease was observed. This fact (the carrier concentration increase as the result of ion irradiation) is unusual for GaAs moderately doped with a shallow impurity and therefore requires an explanation. These results are consistent with the suggestion that in structures with the Mn δ-layer the formation of the Mn impurity band occurs [23]. Moreover, the results can be analyzed under the assumption that the transport properties of the structures are determined by two conduction channels (IB and VB). In this case, the experimental values of the sheet resistance $R_S$ and the Hall coefficient $R_0$ can be expressed as follows [24]:

$$R_S = 1/(ep_{VB}\mu_{VB} + ep_{IB}\mu_{IB}), \quad (1)$$

$$R_0 = \frac{1}{e} \frac{p_{VB}\mu_{VB}^2 + p_{IB}\mu_{IB}^2}{(p_{VB}\mu_{VB} + p_{IB}\mu_{IB})^2}, \quad (2)$$

where $p_{VB}$, $\mu_{VB}$ – the sheet concentration and the mobility of the carriers in the valence band, $p_{IB}$, $\mu_{IB}$ – the sheet concentration and the mobility of the impurity band carriers. Table 1 presents the experimental values of the sheet concentration $p_S$ and the Hall mobility $\mu = R_0/R_S$ for the initial structure at 295 and 90 K (Fig. 2) and $p_S$ and $\mu$ values calculated using (1) and (2) for the proposed values of $p_{VB}$, $\mu_{VB}$, $p_{IB}$, $\mu_{IB}$.

TABLE 1. Experimental, proposed within the framework of two channel model and calculated values of sheet concentration and mobility for the initial structure.

| Experimental $p_S$ and $\mu$ values at 295 K | Experimental $p_S$ and $\mu$ values at 90 K |
|---|---|
| $p_S = 3.1 \times 10^{13}$ cm$^{-2}$, $\mu = 26$ cm$^2$/V·s | $p_S = 1.3 \times 10^{12}$ cm$^{-2}$, $\mu = 423$ cm$^2$/V·s |
| Calculated $p_S$ and $\mu$ values at 295 K ($p_{VB} = 6 \times 10^{11}$ cm$^{-2}$, $p_{IB} = 1 \times 10^{15}$ cm$^{-2}$, $\mu_{VB} = 110$ cm$^2$/V·s, $\mu_{IB} = 0.4$ cm$^2$/V·s) | Calculated $p_S$ and $\mu$ values at 90 K ($p_{VB} = 6 \times 10^{9}$ cm$^{-2}$, $p_{IB} = 1 \times 10^{15}$ cm$^{-2}$, $\mu_{VB} = 5000$ cm$^2$/V·s, $\mu_{IB} = 0.4$ cm$^2$/V·s) |
| $p_S = 2.3 \times 10^{13}$ cm$^{-2}$, $\mu = 20$ cm$^2$/V·s | $p_S = 1.2 \times 10^{12}$ cm$^{-2}$, $\mu = 350$ cm$^2$/V·s |

As the reference value of the sheet concentration $p_{VB}$ at 295 K, the $p_{VB}$ value for the reference structure with the carbon δ-layer was taken ($p_{VB} \sim 10^{12}$ cm$^{-2}$, the structure 7669 in [23]), since this structure has the built-in electric field ($\approx 36$ kV/cm) comparable with that of the structure with Mn δ-layer ($\approx 23$ kV/cm) [23,25]. As the reference values of the hole mobility $\mu_{VB}$ the corresponding experimental values obtained for the other reference structure (uniformly Mn doped GaAs layer) were used. This reference structure was ≈ 1.5 μm-thick layer obtained by MOCVD at 620 °C and doped during the growth process by the laser sputtering of the Mn target. The studies of transport properties in the temperature range of 90 - 290 K revealed that in this MOCVD (Ga,Mn)As the Mn impurity acts as an acceptor with the ionization energy $E_A = 0.1$ eV with no evidence of the impurity band formation. Hence, the carriers are VB holes due to the thermal ionization of the Mn acceptor. The values of the mobility $\mu \approx \mu_{VB}$ were 283 and 12633 cm$^2$/V·s at 290 and 90 K, respectively. On the basis of these values, taking into account that for a heavily manganese doped structure (as in the Mn δ-layer) the mobility should be less than the reference one and the $E_A$ value can be assumed to be ~ 0.07 eV (that leads to the less holes freeze out in the valence band at 90 K [26]), the $p_{VB}$ and $\mu_{VB}$ values were proposed (Table 1). The value of the sheet concentration in the impurity band of $p_{IB} = 10^{15}$ cm$^{-2}$ and the mobility in the impurity band $\mu_{IB} \sim 0.4$ cm$^2$/V·s used to obtain the calculated values of $p_S$ and $\mu$ are reasonable (as it will be shown below, such $p_S$ and $\mu$ values are obtained experimentally after irradiation). The obtained results (Tab. 1) allow us to conclude that the conductivity of the initial structure with the Mn δ-layer is determined by the combination of conductivities in the valence and impurity bands, with the IB carriers concentration being much higher than the VB carriers concentration (even at room temperature). This conclusion is confirmed by the studies of transport properties after irradiation with helium ions (Fig. 3). The generation of the radiation defects leads to the decrease in VB carriers contribution to the total measured Hall concentration and mobility. This is possible both with the decrease in the VB carriers concentration and mobility. Since the measured conductivity is determined by the combination of the VB and IB conductivity the introduction of radiation defects changes the ratio of the contributions from these two conduction channels. It can be expected that the IB conductivity is modified considerably less since the introduction of radiation defects has no influence on the low mobility of IB carriers. The change in the IB carriers concentration after irradiation requires a separate consideration but it can be expected that after irradiation the IB conductivity is changed significantly less than the VB conductivity. The observed concentration increase and mobility decrease after irradiation (Fig. 3) is a consequence of the discussed weakening of the VB conductivity and the predominance of IB conductivity. Table 2 presents the experimental $p_S$ and $\mu$ values at 295 and 90 K for the structure after the irradiation with the fluence of $1 \times 10^{12}$ cm$^{-2}$ (Fig. 3) and the $p_S$ and $\mu$ values calculated using (1) and (2) for the proposed values of $p_{VB}$, $\mu_{VB}$, $p_{IB}$, $\mu_{IB}$. The experimental $p_S$, and $\mu$ values and the used $p_{VB}$, $\mu_{VB}$, $p_{IB}$, $\mu_{IB}$ values indicate that the IB conductivity dominates after irradiation. The contribution from the VB conductivity is greatly weakened but below room temperature it remains appreciable due to the high $\mu_{VB}$ value. Note that the experimental $p_S$ value at 295 K after irradiation is ~

$1 \times 10^{15}$ cm$^{-2}$ (Fig. 3, Table 2). This $p_S$ value indicates that after the introduction of radiation defects the room temperature conductivity is determined by the IB channel, and such $p_S$ value experimentally confirms the correctness of the proposed $p_{IB}$ value (Tables 1 and 2). Note that such experimental $p_S$ value is reasonable. For the width of Mn doped region is about 8 nm (Fig. 1), the sheet concentration ~ $1 \times 10^{15}$ cm$^{-2}$ gives the volume concentration about $1.2 \times 10^{21}$ cm$^{-3}$. The such volume concentration is appropriate for ferromagnetic (Ga,Mn)As. In particular, in the work [27] the volume carrier concentration about $1 \times 10^{21}$ cm$^{-3}$ was found for 100 nm thick (Ga,Mn)As with Mn content about 1.75-3 at. % and similar Curie temperature (~ 30 K).

TABLE 2. Experimental, proposed within the framework of two channel model and calculated values of sheet concentration and mobility for the structure after the irradiation with fluence of $1 \times 10^{12}$ cm$^{-2}$.

| Experimental $p_S$ and $\mu$ values at 295 K | Experimental $p_S$ and $\mu$ values at 90 K |
|---|---|
| $p_S = 9.5 \times 10^{14}$ cm$^{-2}$, $\mu = 0.3$ cm$^2$/V·s | $p_S = 2.9 \times 10^{13}$ cm$^{-2}$, $\mu = 16$ cm$^2$/V·s |
| Calculated $p_S$ and $\mu$ values at 295 K ($p_{VB} = 4 \times 10^{10}$ cm$^{-2}$, $p_{IB} = 1 \times 10^{15}$ cm$^{-2}$, $\mu_{VB} = 15$ cm$^2$/V·s, $\mu_{IB} = 0.4$ cm$^2$/V·s) | Calculated $p_S$ and $\mu$ values at 90 K ($p_{VB} = 4 \times 10^{8}$ cm$^{-2}$, $p_{IB} = 1 \times 10^{15}$ cm$^{-2}$, $\mu_{VB} = 3000$ cm$^2$/V·s, $\mu_{IB} = 0.3$ cm$^2$/V·s) |
| $p_S = 9.5 \times 10^{14}$ cm$^{-2}$, $\mu = 0.4$ cm$^2$/V·s | $p_S = 2.5 \times 10^{13}$ cm$^{-2}$, $\mu = 12$ cm$^2$/V·s |

The present results allow us to conclude that for both the initial structure and the structure after irradiation, the most of charge carriers providing the conductivity of the structure are localized at the allowed states within the Mn impurity band. The more "metallic" character of the $R_S(T)$ dependences after irradiation with the fluence of $1 \times 10^{11}$ and $1 \times 10^{12}$ cm$^{-2}$ relative to the initial structure (Fig. 2) is due to the IB conductivity domination, since the concentration and mobility of carriers localized in IB depends weakly on temperature. After irradiation with the higher fluence of $3 \times 10^{12}$ and $1 \times 10^{13}$ cm$^{-2}$ at temperatures < 100 K the resistance rise more rapidly than for the initial structure (Fig. 2). The above reasoning allow us to conclude that for both the initial structure and after irradiation the total conductivity at temperatures < 100 K is predominantly determined by the IB conductivity, since the VB hole concentration for $E_A$ ~ 0.07 eV should decrease by more than 8 orders of magnitude as the temperature is lowered from 90 to 30 K. Hence, the observed rapid resistance increase with decreasing temperature < 100 K for the fluence of $3 \times 10^{12}$ and $1 \times 10^{13}$ cm$^{-2}$ (Fig. 2) is associated with a modification of the IB conductivity character with increasing the number of radiation defects. In the initial structure the part of Mn atoms are located at substitutional sites (Mn$_{Ga}$) and form the range of allowed energy states in the band gap (impurity band), the part of Mn atoms located at the interstitial sites (Mn$_I$) and act as a double donor centers. In this case, electrons localized on Mn$_I$ can lower their energy by being captured at Mn$_{Ga}$ cites (at that partially fill the impurity band), as a result, some distribution of charged centers in the GaAs matrix occurs. The IB conductivity is determined by the energy width of the impurity band (and the density of states (DOS)), the carrier concentration in the impurity band and its spatial homogeneity in the structure (in the Mn doped region). The introduction of additional electrically active radiation defects modifies the distribution of charged centers. As a result, a change in the impurity band filling, its spatial distribution and DOS should occur. It is known that for heavily doped semiconductors the compensation degree increase leads to the breakdown of the "metallic" conductivity as a result of an increase in the spatial inhomogeneity of the potential relief and the formation of "metallic" regions separated by the relatively high-resistance regions depleted of free carriers [28]. For the structure with the Mn δ-layer the "metallic" regions are regions with the impurity band and the high-resistance regions are regions of the structure in which the impurity band is destroyed after irradiation as a result of the formation of additional charged centers. With the fluence increasing > $1 \times 10^{12}$ cm$^{-2}$ (consequently with the concentration of additional randomly distributed charged centers increasing) the value and number of "metallic" regions with the impurity band decrease and these regions are increasingly isolated from one another. This leads to an activation-type of $R_S(T)$ dependences at $T < 100$ K (Fig. 1, $F = 3 \times 10^{12}$ and $1 \times 10^{13}$ cm$^{-2}$).

### 3.3. Magnetic properties

Figure 4 shows the Hall resistance as a function of magnetic field ($R_H(H)$) at different temperatures for the initial structure and the structure after irradiation with the fluence of $1 \times 10^{12}$ and $1 \times 10^{13}$ cm$^{-2}$. For the initial and irradiated structure the $R_H(H)$ dependences at $T < 50$ K are nonlinear due to the manifestation of the anomalous Hall effect [10]. At 13 K the $R_H(H)$ dependences are hysteretic before and after irradiation (inset in Fig. 4(a)). After irradiation with $F = 1 \times 10^{12}$ cm$^{-2}$ the coercive field ($H_C$) increases from 35 to 160 Oe and after irradiation with $F = 1 \times 10^{13}$ cm$^{-2}$ returns to the approximately initial value (inset in Fig. 4(a)). The hysteresis loop on the $R_H(H)$ curve after irradiation with $F = 1 \times 10^{12}$ cm$^{-2}$ is also observed at 20 K (Fig. 4(b)). Probably $H_C$ increase is associated with the modification of the magnetic anisotropy of the Mn δ-layer. In the initial structure a predominantly in-plane magnetic anisotropy is observed as a consequence of the low thickness of the Mn doped region [13]. The introduction of additional charged centers after irradiation, as discussed above, can lead to a more inhomogeneous spatial distribution of regions with the Mn impurity band ("metallic" regions) that can modify the character of the magnetic anisotropy. The $H_C$ increase after ion irradiation was observed earlier for (Ga,Mn)As, note that it was accompanied by $T_C$ decrease [6,7].

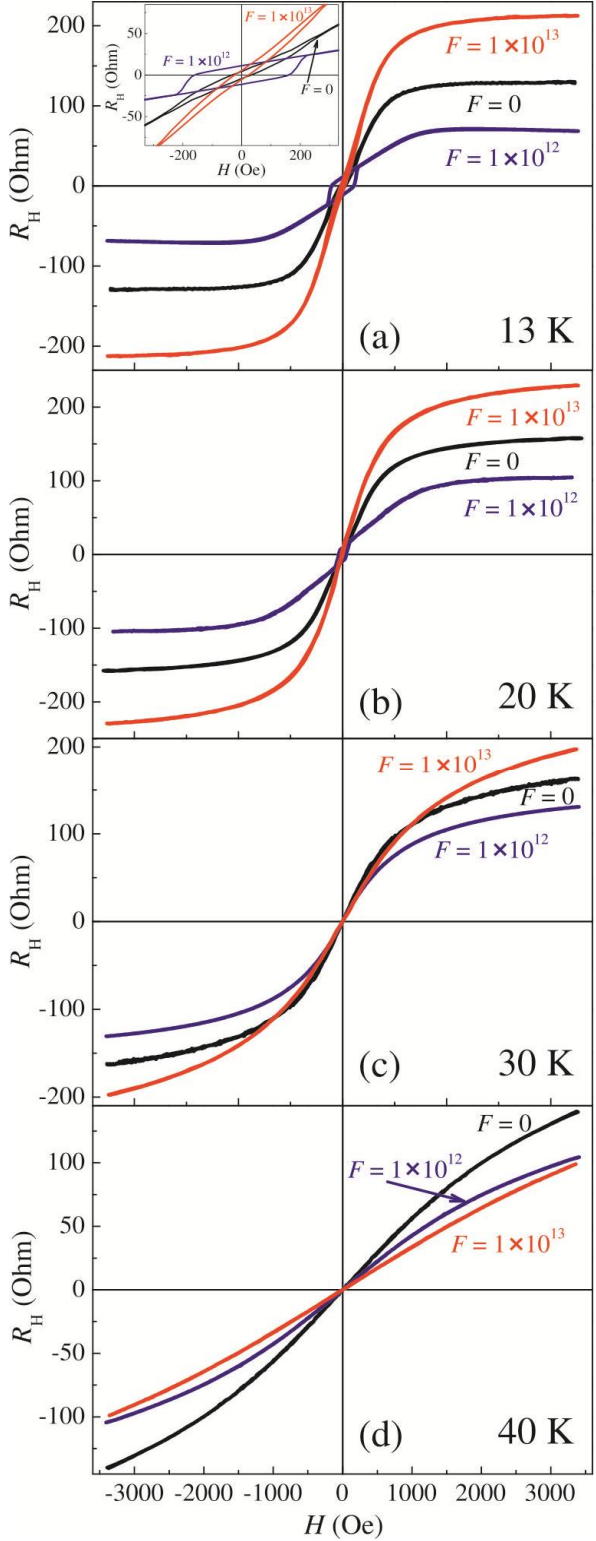

**Fig. 4.** (a)-(d) $R_H(H)$ curves at 13, 20, 30 and 40 K for the initial structure and after irradiation with the fluence of $1 \times 10^{12}$ and $1 \times 10^{13}$ cm$^{-2}$.

Let us consider ion irradiation influence on $T_C$ of the structure under study. The $R_S(T)$ dependence for the initial structure has the hump at $T \approx 30$ K, therefore, the value of $T_C$ can be estimated by the given temperature, which is typical for our Mn δ-doped GaAs structures [12]. After irradiation with the fluence of $1 \times 10^{11}$ and $1 \times 10^{12}$ cm$^{-2}$ the hump on $R_S(T)$ dependences becomes more pronounced. Then after irradiation with the fluence of $3 \times 10^{12}$ and $1 \times 10^{13}$ cm$^{-2}$ the hump becomes less noticeable (Fig. 2). Figure 5 shows the sheet resistance derivatives ($dR_S(T)/dT$) in the temperature range of 14 - 45 K for the structure before and after irradiation. For the initial structure the maximum of the $dR_S(T)/dT$ dependence is observed at ≈ 26 K. It was shown in [29] that for (Ga,Mn)As with the relatively low Curie temperature the maximum of the derivative is observed at $T$ lower than $T_C$ determined from magnetization measurements (about 14% for the case considered in [29]). Taking this into account, the $dR_S(T)/dT$ dependence allow us to estimate $T_C$ also about 30 K for the initial structure. For the fluence of $1 \times 10^{11}$ and $1 \times 10^{12}$ cm$^{-2}$ the peak position does not change (a slight shift toward a higher temperature for the fluence of $1 \times 10^{12}$ cm$^{-2}$ is observed). For the fluence of $3 \times 10^{12}$ and $1 \times 10^{13}$ cm$^{-2}$ a relatively rapid resistance rise at $T < 50$ K leads to the expansion of the derivative peak, but it can be seen that the bend at ≈ 26 K persists (Fig. 5).

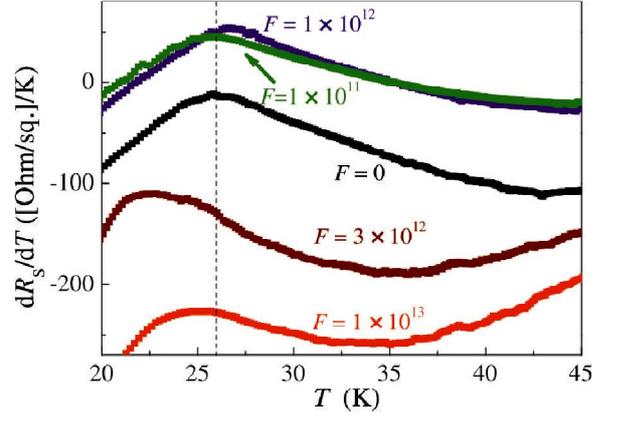

**Fig.5.** $dR_S(T)/dT$ dependences for the structure before and after irradiation.

The $T_C$ can also be estimated by Arrot plots of the $R_H(H)$ curves. Figures 6(a)–6(c) show Arrot plots for the initial structure and after irradiation with the fluence of $1 \times 10^{12}$ and $1 \times 10^{13}$ cm$^{-2}$. It is seen from the plots that the structure both before and after irradiation is in the ferromagnetic state at 30 K and paramagnetic at 40 K. For all used fluences, the sufficient number of ferromagnetic "metallic" regions with $T_C \approx 30$ K preserve in the structure, so that the magnetotransport properties (in particular AHE) are determined by them. These "metallic" regions after irradiation with the fluence of $3 \times 10^{12}$ and $1 \times 10^{13}$ cm$^{-2}$, as discussed previously, are separated by high-resistance paramagnetic (or with lower $T_C$) regions. The Hall effect (as well as the resistance) after irradiation with the relatively high fluence of $1 \times 10^{13}$ cm$^{-2}$ is determined both by "metallic" ferromagnetic (with $T_C \approx 30$ K) and high-resistance paramagnetic regions, which have an influence on the shape of the $R_H(H)$ curves, hence, on the Arrot plots. In general, it can be concluded that for the Mn δ-doped GaAs structure the introduction of radiation defects *does not lead* to the $T_C$ decrease with fluence increasing. The $T_C$ invariability after the introduction of radiation defects was also observed in [4] for 200 nm-thick (Ga,Mn)As and it was explained by the mutual compensation of the $T_C$ decreasing effect with carrier concentration decreasing by the $T_C$ increasing effect due to an increase in disorder. In our opinion, the $T_C$ invariability is a natural consequence of the fact that carriers (holes) in the valence band do not affect the magnetic properties of the structure. For the Mn doped regions in the Mn δ-doped GaAs structure, where after the creation of additional charged centers the Mn impurity band is preserved, the Curie temperature remains unchanged (≈ 30 K).

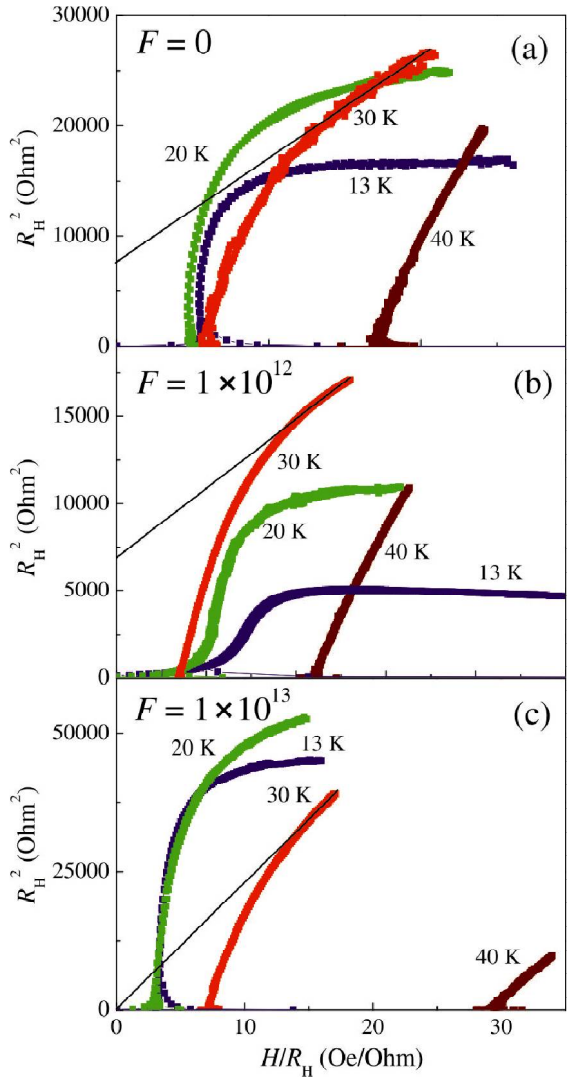

**Fig. 6.** (a) - (c) Arrot plots of $R_H(H)$ curves.

### 4. Disscussion

The transport and magnetic properties of Mn δ-doped GaAs structure before and after irradiation are determined by carriers with energy lying in the range of allowed states within the Mn impurity band. The obtained results agrees with the impurity band picture of magnetism in (Ga,Mn)As (Zener's double exchange model). Both for Zener's double exchange model and for Zener's *p-d* exchange model the exchange interaction between Mn ions are carrier-mediated. In the case of Zener's *p-d* exchange, the carriers are holes at the top of the valence band. For Zener's double-exchange model (impurity band picture), holes are also traditionally considered as carriers providing exchange interaction, but located in the impurity band. In the work [30] using Zener double exchange Hamiltonian for holes in a narrow impurity band it was shown that within an impurity band model, partial compensation promotes the ferromagnetic order in (Ga,Mn)As, with the highest Curie temperature reached for the case of 0.5 holes per substitutional Mn (i.e. for a half-filled impurity band). For a completely empty or completely filled impurity band the ferromagnetism disappears. The problem is traditionally formulated using holes both as intermediaries in the exchange interaction and as charge carriers. At the same time, the used in [30] approach and the Hamiltonian can be applied equally to electrons located at the states within the Mn impurity band. In the limiting case, when only $Mn_{Ga}$ atoms are present in GaAs (and no any donors) the Mn impurity band can form. This is the region of allowed states of finite width in the forbidden band which is separated from the valence band edge by a gap of finite width (or a gap with a small DOS of the band`s tails). In this case, the impurity band can equally be called both completely empty for electrons and completely filled with holes. It is obvious that at zero temperature the conductivity in this system is absent, since the $Mn_{Ga}$ atoms are an array of indistinguishable neutral acceptors. Within the framework of Zener's double-exchange model the ferromagnetism is also absent [30]. The conductivity as well as the exchange interaction between Mn atoms according to Zener's double-exchange model can appear when the impurity band is partially filled with electrons, for example, as a result of partial compensation. In this case, both the charge carriers and the carriers in carrier-mediated ferromagnetism should be *electrons but not holes*, as traditionally considered, including within the impurity band model. The electron density increase in the impurity band (from completely empty) should lead to the increase of the conductivity and the magnitude of the ferromagnetic exchange interaction (in particular $T_C$). When the impurity band is completely filled with electrons the conductivity should disappear again and the disappearance of ferromagnetism is predicted in [30]. In a real structure the case of a completely empty or completely filled Mn impurity band cannot be realized. In real (Ga,Mn)As electrically active donors (in particular $Mn_I$), which lead to a partial filling of the impurity band by the electrons, are always present. Apparently, the independent filling of IB by electrons for IB with a fixed width and DOS (for the modification of transport and magnetic properties) is difficult to realize experimentally.

There is a opinion that the InGaAs quantum well near the Mn δ-layer has a strong influence on the ferromagnetic properties of the structure [31], similar to which was supposed for MBE Mn delta-doped structures with the quantum well of GaAs/AlGas heterojunction [10,11]. Let us examine the possible influence of InGaAs quantum well presence near (a distance of 2 - 10 nm) the Mn δ-layer. In principle, the presence of the InGaAs quantum well can affect the filling degree of the impurity band by electrons, hence, the influence on the Curie temperature can also be expected. Apparently, the QW influence on the IB filling degree is negligible. The sheet concentration of carriers localized in the InGaAs quantum well does not exceed $\sim 1 \times 10^{12}$ cm$^{-2}$ both for $p$- and $n$-type structures [19,32,33]. Such sheet carrier concentration in QW is much smaller than the sheet carrier concentration in IB ($\sim 1 \times 10^{15}$ cm$^{-2}$, the thicknesses of QW and the Mn δ-doped region are comparable). Neither the quantum well composition nor its thickness has a significant influence on the carrier concentration in it (in the direction of concentration increase). Therefore, the InGaAs quantum well presence does not affect the impurity band filling by the carriers and consequently does not affect the ferromagnetic properties of the GaAs structure with the Mn δ-layer.

## 5. Conclusion

The following conclusions can be made based on the presented results. (i) The transport properties of the GaAs structures with the single Mn δ-layer are determined by two parallel conduction channels - the channel associated with hole transport in the valence band and the channel associated with electron transport in the Mn impurity band, separated from the valence band. (ii) The ferromagnetic properties of the GaAs Mn δ-doped structures are determined by the electrons localized at the allowed states within the Mn impurity band. (iii) The ferromagnetic properties of the Mn δ-layer region cannot be affected by the closely located InGaAs quantum well, since the presence of QW has negligible influence on the Mn impurity band filling by electrons.

## Acknowledgements

This study was supported by the Ministry of Education and Science of Russian Federation (Projects No. 8.1751.2017/PCh).